\documentclass[twocolumn,aps,prl]{revtex4}
\usepackage{graphicx}
\usepackage{amsmath}
\usepackage{amssymb}
\usepackage{wasysym}

\begin{document}

\title{Aspect ratio dependence of charge transport in turbulent electroconvection}

\author{Peichun Tsai,$^{1}$ Zahir A. Daya,$^{1,2}$ and Stephen W.
Morris$^{1}$}
\address{$^{1}$Department of Physics, University of
Toronto, 60 St. George St., Toronto, Ontario, Canada M5S 1A7\\
$^{2}$Center for Nonlinear Studies and Condensed Matter \& Thermal
Physics Group, MS-B258,\\ 
Los Alamos National Laboratory, Los Alamos, NM 87545}

\date{\today}

\begin{abstract}
We present measurements of the normalized charge transport or Nusselt
number $\rm Nu$ as a function of the aspect ratio $\Gamma$ for
turbulent convection in an electrically driven film. In analogy with
turbulent Rayleigh-B{\'e}nard convection, we develop the relevant
theoretical framework in which we discuss the local-power-law-scaling of $\rm Nu$ with
a dimensionless electrical forcing parameter $\cal R$. For these
experiments where $10^4 \lesssim {\cal R} \lesssim 2 \times 10^5$ we
find that ${\rm Nu} \sim {\rm F}(\Gamma){\cal R}^\gamma$ with $\gamma$ either $
= 0.26~(\pm 0.02)$ or $0.20~(\pm 0.03)$, in excellent agreement with the theoretical predictions of $\gamma = 1/4$ and $1/5$. Our measurements of the aspect ratio-dependence of
${\rm Nu}$ for $0.3 \leq \Gamma \leq 17$ compares favorably with the
function ${\rm F}(\Gamma)$ from the scaling theory. 

\end{abstract}

\maketitle

Heat transport has for many years been the cornerstone in the study
of turbulent Rayleigh-B{\'e}nard convection (RBC)~\cite{kadanoff_01}. Over
the past three years there has been remarkable
experimental~\cite{ahlers_00,ahlers_side_00,niemela_00,ahlers_01,xia_lam_zhou_02, ahlers_03,niemela_03} and
theoretical~\cite{GL_00,GL_01,GL_sidewall_03} progress in
characterizing the properties and  mechanisms of heat transfer in
fluids heated from below. So far, 
experiments and theory are relevant only to approximately unit aspect
ratio systems {\em i.e.} geometries that have comparable lateral and
vertical dimensions. Turbulent convection flows in nature, however,
are laterally extended and the transport of heat across a
layer of fluid has been inadequately studied as a function of the
aspect ratio $\Gamma = {\rm lateral~dimension}/{\rm
  vertical~dimension}$. In particular, experiments at $\Gamma > 2$
have seldom been performed primarily because it is extremely difficult
to achieve the strong forcing that is readily reached in $\Gamma \approx 1$
containers. 

In this Letter, we present a study of an analogous electrically driven convecting system
which allows a broad range of $\Gamma$ to be explored at moderate levels of forcing.  We show that measurements of the normalized charge
transport $\rm Nu$ varies with the aspect ratio $\Gamma$ consistent
with a function $\rm F$ given by the scaling theory. The strong dependence on $\Gamma$ especially for $\Gamma \approx 1$ implies that experiment and theories relevant to
this regime are restrictive. The data and
the function $\rm F$ become increasingly independent of $\Gamma$ for
$\Gamma \gg 1$.  This observation highlights the importance of developing experiments and
theory for laterally extended systems where it appears that universal
behavior, often lacking in $\Gamma \approx 1$ systems, may be
restored~\cite{daya_ecke_01}. Our data also reveal that ${\rm Nu} \sim {\cal
  R}^{\gamma}$ with $\gamma = 0.26$ and $0.20$ in agreement with a local-power-law-scaling theory developed in a manner identical to the Grossmann-Lohse (GL) model for
turbulent RBC~\cite{GL_00,GL_01}. 

\begin{figure}
\includegraphics[height=3cm]{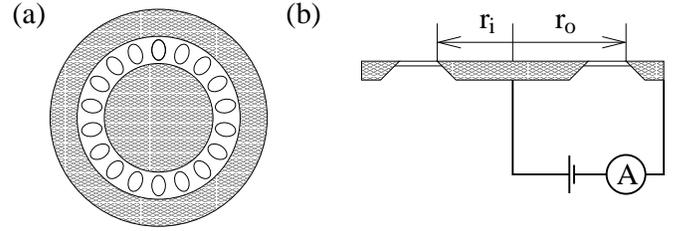}
\caption{\label{schematic}
Schematics of the experiment: (a) top and (b) side.}
\vspace{-0.40cm}
\end{figure}

In previous work, it has been 
demonstrated that an electrically conducting,
freely suspended liquid crystal film between parallel wires could be driven to
convect when a sufficiently large potential drop was applied across
its edges~\cite{morris_90,mao_97,daya_97,dey_97,daya_98,daya_99,daya_thesis_99,daya_01,daya_02}.
Experiments~\cite{daya_98,daya_thesis_99,daya_01,daya_02} and theory~\cite{daya_99,daya_thesis_99} were
then extended to the naturally periodic geometry of an annular film
shown schematically in Fig.~\ref{schematic}. Analogous to
RBC where an inverted mass density distribution is unstable to buoyant
forcing, the thin film has an inverted surface charge density distribution
that is unstable to electrical forcing. In this system, an
initial bifurcation to convection rolls occurs when the applied
voltage $V$ exceeds a critical voltage $V_c$, corresponding to the 
critical temperature difference in RBC. Secondary bifurcations occur at higher
forcing and result in changing the number of counter-rotating vortex
pairs that are arranged around the annulus. At much higher forcing the
fluid becomes turbulent but retains the large scale structure of the
rolls or convection cells. 

Electroconvecting smectic films have several advantages and disadvantages relative
to traditional RBC experiments. In turbulent RBC experiments, the
accurate measurement of heat transport requires detailed accounting
of the heat conducted through the sidewalls~\cite{ahlers_side_00}. The
annular geometry of the smectic film (see Fig.~\ref{schematic}a) is free of lateral
boundaries and consequently without sidewall losses, which facilitates
precise measurement of the charge transport between the inner and
outer electrodes.  The film is an annular disk of
width $d=r_o-r_i \sim 5$ mm and thickness $s \sim 0.1~\mu$m. Here $r_i~(r_o)$
are the radii of the inner (outer) electrodes that support the film
(see Fig.~\ref{schematic}b). These radii can be varied so as to achieve different 
values of $\Gamma$.  The tiny size of the film, which contains many orders of magnitude
less working fluid than required for an RBC experiment, results in data acquisition timescales
of minutes, rather than days.  On the other hand, the smectic film is delicate and while many films survive vigorous forcing without thickness change, some do suffer sudden thickness variations.  Since $V_c$ and other properties of the flows depend on thickness, these data must be discarded.  The dc electrical forcing also results in conductivity drifts in the films, which result in systematic uncertainties. The experiments we describe here were carried out at atmospheric pressure, which increases the film stability.  But air drag, which is known to produce a quantitative but not qualitative change to the dissipation, is not properly accounted for in the theory.  Nevertheless, notwithstanding these caveats, we are able to efficiently  explore the turbulent scaling regime over a broad range of $\Gamma$, complementing  RBC experiments. 

\begin{figure}
\includegraphics[height=6cm]{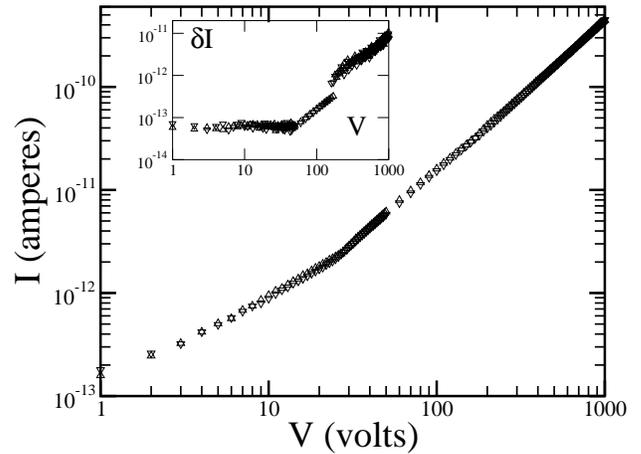}
\caption{\label{iv}
  A representative current-voltage curve for an annular film
  showing the onset of electroconvection. Data obtained for
  increasing (decreasing) voltages are shown in $\bigtriangleup
  (\bigtriangledown)$. The inset shows the rms fluctuations of the
  current {\it vs.} the applied voltage. The transition to turbulent
  flow is marked by a sudden large increase in the fluctuations. Here
 $\Gamma = 6.60 \pm 0.02$ and ${\cal P}=36 \pm 1$. 
 }
\vspace{-0.40cm}
\end{figure}

Our experiment consists of a temperature controlled
 film of octylcyanobiphenyl (8CB), a smectic-A liquid crystal,
suspended between two concentric gold-plated electrodes.
In the smectic-A phase, the film flows as an
isotropic, incompressible and Newtonian fluid in the plane of the
film. The flow is strictly two-dimensional. The film is driven to
electroconvect by a dc voltage applied to its inner edge while holding
the outer edge at ground potential. The annular assembly is enclosed
in a Faraday cage. An experiment consists of imposing a voltage $V$ to
the inner electrode and measuring the current $I$ transported through
the film with a sensitive electrometer. The applied voltage was varied between $0$ and $1000$ volts in a sequence of small incremental and decremental steps resulting in
a current-voltage characteristic. Further details of the apparatus and
procedure can be found in Refs.~\cite{daya_98,daya_99,daya_thesis_99,daya_01,daya_02}. 
Figure~\ref{iv} shows a
representative current-voltage (IV) characteristic. For $V < V_c$ the fluid
is quiescent with the current sustained by ohmic conduction. When $V >
V_c$ the fluid is organized in counter-rotating vortices and additional 
charge is transferred by convection; this is seen by the
increase in slope of the IV-characteristic. At higher voltages, the
fluid becomes turbulent and the transition is marked by a sudden
increase in the rms fluctuations of the current, as shown in the inset
of Fig.~\ref{iv}. 

A film is a 2D annular sheet. The film has thickness/width $s/d\sim
10^{-4}$. Geometrically, a film is described by its radius ratio
$\alpha = r_i/r_o$ and the aspect ratio $\Gamma_r =
\pi(r_i+r_o)/(r_o-r_i)$, which is the ratio of the mid-radius
circumference to the film width. So defined, $\pi < \Gamma_r < \infty$.  To make correspondence with the conventional aspect ratio $\Gamma$ for RBC, we define $\Gamma = \Gamma_r - \pi$ for the annulus.  Two other dimensionless parameters
describe the experimental system: ${\cal R}={\epsilon_0^2 V^2}/{\sigma
  \eta s^2}$ and ${\cal P}={\epsilon_0\eta}/{\rho \sigma s d}$. Here
${\cal R}$ is the control parameter and is a measure of the external
electrical forcing and ${\cal P}$, the Prandtl-like parameter, is the
ratio of the time scales of electrical and viscous dissipation
processes in the film. In the above the fluid density, molecular
viscosity, and conductivity are denoted by $\rho$, $\eta$, and
$\sigma$, and $\epsilon_0$ is the permittivity of free space. 

\begin{figure}
\includegraphics[height=6cm]{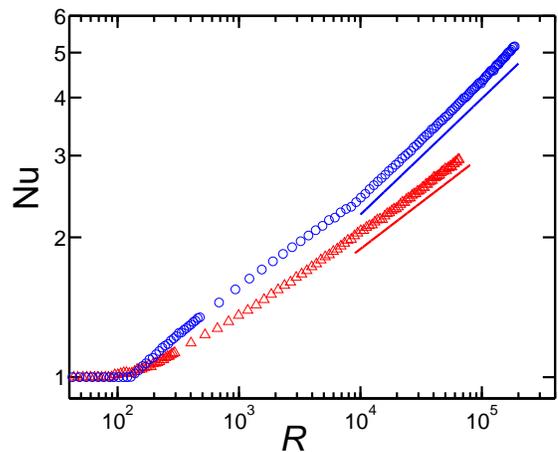}
\caption{
\label{nusselt} Plots of ${\rm Nu}$ vs ${\cal R}$ for $\Gamma = 0.33 \pm 0.01 (\triangle)$ and $6.60 \pm 0.02 (\circ)$. For these ${\cal P}=8.8 \pm 0.3 $ and $36 \pm 1$ respectively. The solid reference lines have slopes of $1/5$ and $1/4$ respectively.
}
\vspace{-0.40cm}
\end{figure}

We normalize the measured total electric current $I$ by the portion due
to conduction $I_{cond}=cV$ where $c$ is the ohmic conductance of the
film obtained from the IV-characteristic when the film is not convecting, {\em  i.e.} when $V < V_c$. We define the normalized charge transport or
Nusselt number ${\rm Nu}= I/I_{cond}$. Figure~\ref{nusselt} shows
representative sets of ${\rm Nu}$ vs ${\cal R}$ data at 
$\Gamma= 0.33 \pm 0.01$ and $6.60 \pm 0.02$.
We find that ${\rm Nu} \sim {\cal R}^\gamma$ for $ 10^4
\leq {\cal R} \leq 2 \times 10^5$ with either $\gamma = 0.26 \pm 0.02$ or $\gamma= 0.20 \pm 0.03$, depending on the size of ${\cal P}$. The
error bar is obtained from the variation in the best-fit value of
$\gamma$ over $26$ data sets for all four values of $\Gamma$.

We developed a theoretical model describing turbulent electroconvection
by borrowing from turbulent RBC. The
equations of motion that describe electroconvection consist of the
incompressible Navier-Stokes equation supplemented with an electrical
body force, the charge conservation equation that accounts for ohmic
conduction and advection of charge and Maxwell's equation that
relates the surface charge density to the electric potential;
\begin{eqnarray}
{\partial}_t {\bf u} + {\bf u} \cdot \nabla {\bf u} &=& -\nabla \frac{p}{\rho} +
\nu \nabla^2 {\bf u} - \frac{q}{\rho} \nabla \psi, \label{ns} \\
{\partial}_t q + {\bf u} \cdot \nabla q &=& \sigma {\nabla}^2 \psi, \label{ct} \\
q &=& -2 \epsilon_0 \partial_z \psi\, .
\end{eqnarray}
These equations are constrained by the no-slip and applied electric
potential boundary conditions. In the above equations $\nabla$, ${\bf u}$, $p$
and $\nu$ are the 2D gradient, velocity, pressure and kinematic viscosity
respectively. The electric potential $\psi$ is three dimensional and extends outside the film,
where $({\nabla}^2 + \partial^2_z)\psi=0$.  The Maxwell equation
relates the perpendicular gradient of $\psi$ to the surface charge $q$.  The factor of 2
arises from the film's two free surfaces.  The relation between $\psi$ and $q$ is thus nonlocal
and somewhat complex.  It can be considerably simplified by making a local approximation, setting  $q=2\beta\psi$, where $\beta$ is a certain constant. This
local approximation has been shown to be adequate in describing the onset of electroconvection. See Ref.~\cite{daya_99} for a
detailed discussion of the theoretical model.

The Eqns.~\ref{ns},~\ref{ct} are similar to the 
Boussinesq equations for turbulent RBC.  In particular, in the local approximation, there is an almost precise correspondence between the temperature and the electric potential. 
The periodicity of the annular geometry and thereby of
the velocity and electric potential allow for exact relations for the
globally-averaged kinetic and electric dissipations. Denoting averages
over the system volume by $\langle \cdot \cdot \cdot  \rangle$ we find the following for the
kinetic dissipation ${\epsilon}_{\bf u} \equiv \langle \nu (\nabla {\bf
  u})^2 \rangle $ and for the electric dissipation ${\epsilon}_\psi \equiv
\langle \sigma (\nabla \psi)^2 \rangle$:
\begin{eqnarray}
{\epsilon}_{\bf u} &=& \frac{\nu^3 {\cal R} {\cal P}^{-2} ({\rm Nu} - 1)}{\ln(1/\alpha)(r_o^2-r_i^2)(r_o-r_i)^2} \,,\\
{\epsilon}_\psi &=& \frac{2\sigma V^2 {\rm Nu}}{\ln(1/\alpha)(r_o^2-r_i^2)}\,.
\end{eqnarray}
These relations are similar to those for the kinetic and thermal
dissipation in RBC presented in
Refs.~\cite{GL_00,GL_01,shr_sig_90_94}. Following the GL theory, we decompose the kinetic and
electric dissipations into bulk and boundary layer parts:
$\epsilon_{{\bf u},\psi} = \epsilon_{{\bf
    u},\psi}^{BULK}+\epsilon_{{\bf u},\psi}^{BL}$. Considering the
various combinations of the dominant contributors to the total
dissipation ({\it e.g.} BL-$\psi$ and BULK-${\bf u}$ {\it etc.}) we can
determine the corresponding local power-law scalings for each of the
regimes. We assume that the turbulent
flow is comprised of convection cells that tile the annulus as shown in
Fig.~\ref{schematic}a and that each cell is roughly square {\em
  i.e.} it has the same transverse and lateral dimension. The large
scale circulation is then the vortex that defines a cell. Assuming
laminar boundary layer scaling we can then show that 
\begin{eqnarray}
{\rm Nu} &\sim& {\rm F}(\Gamma) {\cal R}^{\gamma} {\cal P}^{\delta}\,,
\hspace{0.25cm} {\rm Re} \sim {\cal R}^{\gamma_\ast}{\cal P}^{\delta_\ast}\,, ~{\rm and} \\
{\rm F}(\Gamma)&=& \frac{\Gamma + \pi}{\pi} \ln\Biggl(\frac{\Gamma+2\pi}{\Gamma}\Biggr)\,.
\end{eqnarray}

As might be expected, we find the same set of exponents
$\gamma,\delta,\gamma_\ast,\delta_\ast$ that appear in the GL
theory. Here, ${\rm Re}$ is the Reynolds number of the large scale
circulation. 

In the relatively small ${\cal R}$ regime, the total dissipation is dominated by the BL-contributions and the theory predicts $\gamma=1/4$ and $\delta=1/8$.  In a neighboring regime where the dissipation is primarily $\epsilon_{{\bf u}}^{BULK}$ and $\epsilon_{\psi}^{BL}$, the theory predicts $\gamma =\delta = 1/5$.
Our measured $\gamma$ exponents, as shown in Figure~\ref{nusselt} are in reasonable agreement with the theoretical predictions from the GL model, if we suppose that we traverse the appropriate regimes as ${\cal P}$ varies.
A detailed validation of the theory would depend crucially on either directly demonstrating the correctness of
the assumptions about the dominant contributions to the dissipation or on 
systematically showing that the right local power-law scalings are
indeed obtained as one varies ${\cal P}$ and ${\cal R}$.  Our experiments currently span only relatively low ${\cal R} < 2\times 10^5$ and sparsely cover the rather wide range $5 < {\cal P} < 250$. Many more experiments will be needed to adequately test the GL theory for turbulent electroconvection.

In the theoretical treatment described above, we have taken into
account the dependence on the aspect ratio of the system, a
consideration that was missing in the GL theory which treated only
$\Gamma \approx 1$
systems~\cite{GL_00,GL_01,GL_sidewall_03}.
We find that the charge transport is modified by a $\Gamma$-dependent factor ${\rm  F}$($\Gamma$). Since the large scale circulation is local to each
convection vortex, ${\rm Re}$ is independent of $\Gamma$. Unlike Ref.~\cite{shr_sig_90_94}, 
we find the aspect-ratio
dependence is not a power-law scaling but rather a
function of the finite annular geometry. We can make a direct comparison with previous turbulent RBC experiments by making a correspondence between the different cell geometries, using $\Gamma = \Gamma_r - \pi$ as described above. We plot $k{\rm F}$($\Gamma$) {\it vs.} $\Gamma$, with the constant  $k=\pi/((\pi+1)\ln(2\pi+1))=0.382$.  Following Ref.~\cite{ahlers_00}, we choose this normalization so that $k{\rm F}(\Gamma=1)=1$. Then $k{\rm F}
\rightarrow 0.764$ in the $\Gamma \rightarrow \infty$ limit. The
function $k{\rm F}$ decreases monotonically with $\Gamma$ with the
greatest variation for $\Gamma < 2$, and is within $2\%$ of the
limiting value for $\Gamma > 7$. 

\begin{figure}
\includegraphics[height=6cm]{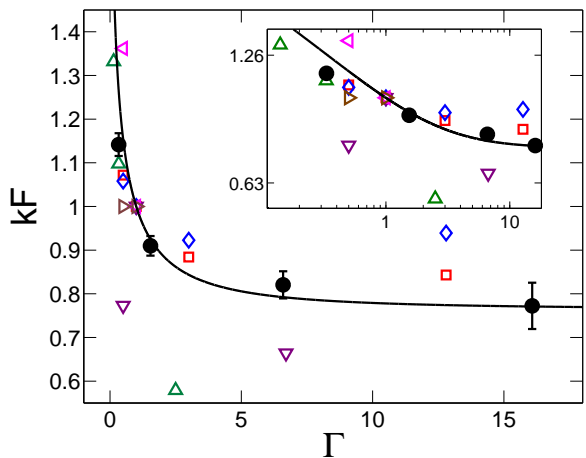}
\caption{\label{aspect}
  The function $k{\rm F}$ vs $\Gamma $. $k{\rm F}(\Gamma)$
  is normalized to unity at $\Gamma=1$. The solid circles ($\bullet$) are
  from the present work. Data from RBC experiments in Acetone, with ($\Diamond$) and without ($\Box$) sidewall correction are from  Refs.~\protect{\cite{ahlers_side_00}} and \protect{\cite{ahlers_00}}.   Also shown are data in Air ($\bigtriangleup$), from Ref.~\protect{\cite{threlfall_74}}, Helium ($\bigtriangledown$) from Ref.~\protect{\cite{wu_libchaber_92}} and ($\triangleleft$) from Refs.~\protect{\cite{niemela_00,niemela_03}} and water ($\triangleright$) from Refs.~\protect{\cite{HKgroup_96}}. The inset shows the same data on a logarithmic scale.
  }
\vspace{-0.40cm}
\end{figure}

Our experimental data span the range $0.3 \leq \Gamma \leq 17$. From power-law fits to ${\rm Nu}$ {\it vs.} ${\cal R}$ data we have extracted the exponents $\gamma$. To determine ${\rm F} (\Gamma)$ we then divide ${\rm Nu}$ by ${\cal R}^{\gamma} {\cal P}^{\delta}$.
Because our data extend over a rather wide range of ${\cal P}$, we expect to cross regimes with differing $\gamma$ and $\delta$\cite{GL_00,GL_01}. To extract the aspect ratio dependence alone, we used the fitted value of $\gamma$, and where $\gamma \approx 0.25~(0.20)$, we used the GL-theory prediction of $\delta = 1/8~(1/5)$. Using one free parameter for all the data, we again scaled these results so that ${\rm F}(\Gamma=1)=1$. Our data for $4$ different $\Gamma$ are in reasonable quantitative agreement with the theoretical function ${\rm F}$, as shown in Fig.~\ref{aspect}. Each data point is an average over $4-10$ runs. The error estimates are representative of the scatter in ${\rm F}$ over these data. In particular, the largest contribution to the error at $\Gamma = 16$ is systematic and arises from the ambiguity in the correct exponent for ${\cal P}$.

Data from several turbulent RBC experiments\cite{ahlers_00,ahlers_side_00,niemela_00,niemela_03,HKgroup_96}
 for $4$ values of $\Gamma$ are also broadly in agreement with the function ${\rm F}$, in spite of the difference in geometry and the higher range of Rayleigh numbers. Earlier RBC experiments\cite{threlfall_74,wu_libchaber_92} used gases as the working fluid and deviate significantly from ${\rm F}$. 

Several important questions are raised by our results on the aspect
ratio dependence of charge transport in turbulent electroconvection and, by extension, that of heat transport in turbulent RBC. The strong
dependence of ${\rm F}$ on $\Gamma$ for $0 \leq \Gamma
\lesssim 7$, suggests that the dynamics of turbulent
convection is dependent on the system's lateral extent. That the
charge or heat transport, a global property, is sensitive to the
 aspect ratio further suggests that local properties such as
temperature and charge fluctuations may be even more strongly
dependent on $\Gamma$. Experiments in turbulent RBC clearly show
that while heat transport is insensitive to cell geometry at fixed
$\Gamma \approx 1$, the fluctuations are strongly affected by the shape of
the lateral boundary~\cite{daya_ecke_01}. The relative independence of
${\rm F}$ on $\Gamma$ for large aspect ratio, $\Gamma
\gtrsim 7$ suggests that the normalized charge and heat transport
approach a universal value in laterally extended systems.  We conjecture
 that fluctuations in the interior may also become
universal in large aspect ratio systems. 

Our results emphasize the importance of extending convective turbulence experiments to
larger aspect ratios. For turbulent smectic electroconvection, we would like
to make a systematic study of the scaling of ${\rm Nu}$ in the parameter space of ${\cal
  R}$ and ${\cal P}$.  ${\cal P}$ can be varied in principle by changing the thickness and/or
width of the film.   We would also like to study
long time series of ${\rm Nu}$ in order to infer the characteristic
size and duration of the fluctuations. It is a challenge to extend the forcing
parameter ${\cal R}$ to values much beyond $10^6$, since this requires applying large
voltages across the film that may result in dielectric breakdown. A
first step would be to better control the electrochemistry of the liquid
crystal film to reduce drifts and push down the critical voltage $V_c$.  
It would also be interesting to extend our 
previous studies of electroconvection under
shear~\cite{daya_98,daya_99,daya_thesis_99,daya_01,daya_02} to the turbulent
regime.   Finally, we would like to develop local probes
 to study the fluctuations of the velocity and electric
potential over the film.

We thank G. Ahlers, E. Ben-Naim, R. E. Ecke and E. Titi for helpful discussions and constructive comments. This research was supported by the Canadian NSERC and the U.S. DOE 
(W-7405-ENG-36).


\vspace{-0.5cm}
\begin{references}

\bibitem{kadanoff_01} L. P. Kadanoff, Physics Today, {\bf 54}, 8-34
  (2001).


\bibitem{ahlers_00} X. Xu, K. M. S. Bajaj, and G. Ahlers, Phys. Rev. Lett.
{\bf 84}, 4357 (2000).

\bibitem{ahlers_side_00} G. Ahlers, Phys. Rev. E {\bf 63}, 015303(R)
  (2000).

\bibitem{niemela_00} J. J. Niemala, L. Skrbek, K. R. Sreenivasan, and
   R. J. Donnelly, Nature (London) {\bf 404}, 837~(2000).

\bibitem{ahlers_01} G. Ahlers and X. Xu, Phys. Rev. Lett. {\bf 86},
   3320 (2001). 

\bibitem{xia_lam_zhou_02} K.-Q. Xia, S. Lam, and S.-Q Zhou,
   Phys. Rev. Lett. {\bf 88} 064501 (2002).

\bibitem{ahlers_03} A. Nikolaenko and G. Ahlers, Phys. Rev. Lett. {\bf 91}, 084501 (2003).

\bibitem{niemela_03} J. J. Niemela and K. R. Sreenivasan, J. Fluid Mech. {\bf 481}, 355 (2003).

\bibitem{GL_00} S. Grossman and D. Lohse, J. Fluid
   Mech. {\bf 407}, 27~(2000). 

\bibitem{GL_01} S. Grossman and D. Lohse, Phys. Rev. Lett.
{\bf 86}, 3316 (2001).

\bibitem{GL_sidewall_03} S. Grossman and D. Lohse, J. Fluid Mech., {\bf 486}, 105 (2003).

\bibitem{daya_ecke_01} Z. A. Daya and R. E. Ecke, Phys. Rev. Lett. {\bf
    87}, 184501 (2001). 



\bibitem{morris_90} S. W. Morris, J. R. de Bruyn, and A. D. May,
  Phys. Rev. Lett. {\bf 65}, 2378 (1990).

\bibitem{mao_97} S. S. Mao, J. R. de Bruyn, and S. W. Morris, Physica
  A {\bf 239}, 189 (1997).

\bibitem{daya_97} Z. A. Daya, S. W. Morris, and J. R. de Bruyn,
  Phys. Rev. E {\bf 55}, 2682 (1997).

\bibitem{dey_97} V. B. Deyirmenjian, Z. A. Daya, and
  S. W. Morris, Phys. Rev. E, {\bf 56}, 1706 (1997).

\bibitem{daya_98} Z. A. Daya, V. B. Deyirmenjian, S. W. Morris,
and J. R. de Bruyn, Phys. Rev. Lett. {\bf 80} 964 (1998).

\bibitem{daya_99} Z. A. Daya, V. B. Deyirmenjian, and S. W. Morris,
Phys. Fluids {\bf 11}, 3613 (1999).

\bibitem{daya_thesis_99} Z. A. Daya, Ph.D. thesis, unpublished (1999).

\bibitem{daya_01} Z. A. Daya, V. B. Deyirmenjian, and S. W. Morris, Phys. 
Rev. E {\bf 64}, 036212 (2001).

\bibitem{daya_02}  Z. A. Daya, V. B. Deyirmenjian, and S. W. Morris, Phys. 
Rev. E {\bf 66}, 015201(R) (2002).

\bibitem{shr_sig_90_94} B. I. Shraiman and E. D. Siggia, Phys. Rev. A {\bf 42},
   3650~(1990), and E. D. Siggia, Annu. Rev. Fluid Mech. {\bf 26},
  137~(1994).

\bibitem{HKgroup_96} Y. Shen, P. Tong and K.-Q. Xia, Phys. Rev. Lett., {\bf 76}, 908 (1996).

\bibitem{threlfall_74} D. C. Threlfall, Ph.D. thesis, unpublished (1974).

\bibitem{wu_libchaber_92} X-Z Wu and A. Libchaber, Phys. Rev. A {\bf 45}, 842 (1992).



\end{references}
\end{document}